# MUSICAL INSTRUMENT SOUND CLASSIFICATION WITH DEEP CONVOLUTIONAL NEURAL NETWORK USING FEATURE FUSION APPROACH

Taejin Park and Taejin Lee

Electronics and Telecommunications Research Institute (ETRI), Republic of Korea

## Abstract

A new musical instrument classification method using convolutional neural networks (CNNs) is presented in this paper. Unlike the traditional methods, we investigated a scheme for classifying musical instruments using the learned features from CNNs. To create the learned features from CNNs, we not only used a conventional spectrogram image, but also proposed multiresolution recurrence plots (MRPs) that contain the phase information of a raw input signal. Consequently, we fed the characteristic timbre of the particular instrument into a neural network, which cannot be extracted using a phase-blinded representations such as a spectrogram. By combining our proposed MRPs and spectrogram images with a multi-column network, the performance of our proposed classifier system improves over a system that uses only a spectrogram. Furthermore, the proposed classifier also outperforms the baseline result from traditional handcrafted features and classifiers.

*Index Terms*— Convolutional Neural Networks, Multiresolution Recurrence Plots, Musical instrument classification, Music information retrieval

## 1. INTRODUCTION

As a part of music information retrieval and music data analysis, musical instrument classification is one of the most crucial tasks in obtaining high-level information regarding a music signal. For the past couple of decades, one of the biggest issues in classifying musical instruments has been selecting the best feature set for the given classification task [1]. Unlike automatic speech recognition (ASR), which mainly employs Mel-frequency cepstral coefficients (MFCCs), instrument classification uses not only MFCCs, but also spectral [2], temporal [3] and timbral texture [4] features, among others. However, a classification system using these handcrafted feature sets requires a feature processing such as a reduction in the number of dimensions because of its internal redundancy [5, 6]. This feature processing largely determines the performance of the classifier, and a significant amount of effort is required to collect and optimize all of the useful features. However, the recent developments in convolutional neural networks (CNNs) [7] have drastically improved the method for extracting the features from a raw signal, achieving a remarkable image [8] and speech recognition [9] performance without the use of any handcrafted features.

The most common approach to applying CNNs to an audio-recognition task is employing a spectrogram image as the feeding data [9, 10]. On the other hand, there have been a few attempts at feeding a raw time-series audio signal into a neural network for music [11] and speech [12] signals. However, in this paper, we propose another method for feeding the phase information into a neural network. Because a time-series is a sequence of one-dimensional data, the filter (kernel) used for the time-series in CNNs should also be one-dimensional. This restriction has given rise to a limitation when applying a filter convolution because information in a certain temporal location can only be convoluted a single time; in contrast, a spectrogram image, which is two-dimensional, can be analyzed using a two-dimensional filter multiple times per single temporal location, which provides more dimensions for an analysis. Nevertheless, the spectrogram loses the phase information because it computes the magnitude of the Fourier transform coefficient. To overcome this issue, we introduce the concept of multiresolution recurrence plots (MRPs). Using MRPs, time-series data can be analyzed in a two-dimensional space without losing any phase information. Combined with a spectrogram image, MRPs are able to provide additional characteristics from a given time-series signal that can work complementarily with a spectrogram image.

The remainder of this paper is organized as follows. In section 2, we introduce the proposed MRPs. In section 3, we describe the proposed structure of the CNNs and our data feeding technique. In section

4, we evaluate our proposed method through instrument recognition tasks. Finally, in section 5, we summarize the main findings of our work.

## 2. MULTIRESOLUTION RECURRENT PLOTS

A recurrence plot (RP) is a widely used analysis technique that visualizes a square matrix, where each element in the matrix corresponds to the distance between two phase-trajectories in the time space [13]. An RP is an effective way to visualize how a phase trajectory revisits the same area in a given phase space, which can also be described as a recurrence pattern. An RP can be simply computed through the following equation:

$$R(i, j) = |x[i] - x[j]|, \qquad (1)$$

where $x[t]$ is a time-series input. RPs have been used to analyze various time-series data [14], including applications such as medical data [15].

However, a conventional RP is not applicable to the sound signal of a musical instrument because musical instruments have a very wide frequency range, which requires a huge RP image for containing all of the phase information regarding the lowest (20 Hz) to highest (4 kHz) frequencies. Therefore, we employ MRPs, which contain a different temporal block size and resolution for each layer.

**2.1. Multilayer analysis of MRPs**

Each MRP for each layer of the MRPs contains a time-series sample block size of $2^n$ in length (where n is an integer). Although each layer contains a different sized time-series signal block, the final MRP image is always reduced to a 32 x 32 matrix. Therefore, we finally have multiple MRP matrices, each layer of which contains a different time span, as shown in figure 1. A lower layer, which contains a longer temporal span, is able to reveal lower-frequency phase information, whereas a higher layer, which contains a shorter temporal span, is able to reveal higher-frequency phase information, and thus the whole layer set becomes a set of multiresolution images. However, the limitation of MRPs is that they are unable to show the exact temporal envelope for each frequency in the way a spectrogram is able to do. Moreover, to include full temporal data from a time series, the MRPs should include quadruple the amount of data for every layer, from the lowest layer to the highest. As a result, to reduce the data size and maintain the balance between layers, we only take the first MRP for each layer while omitting the rest of the temporal data.

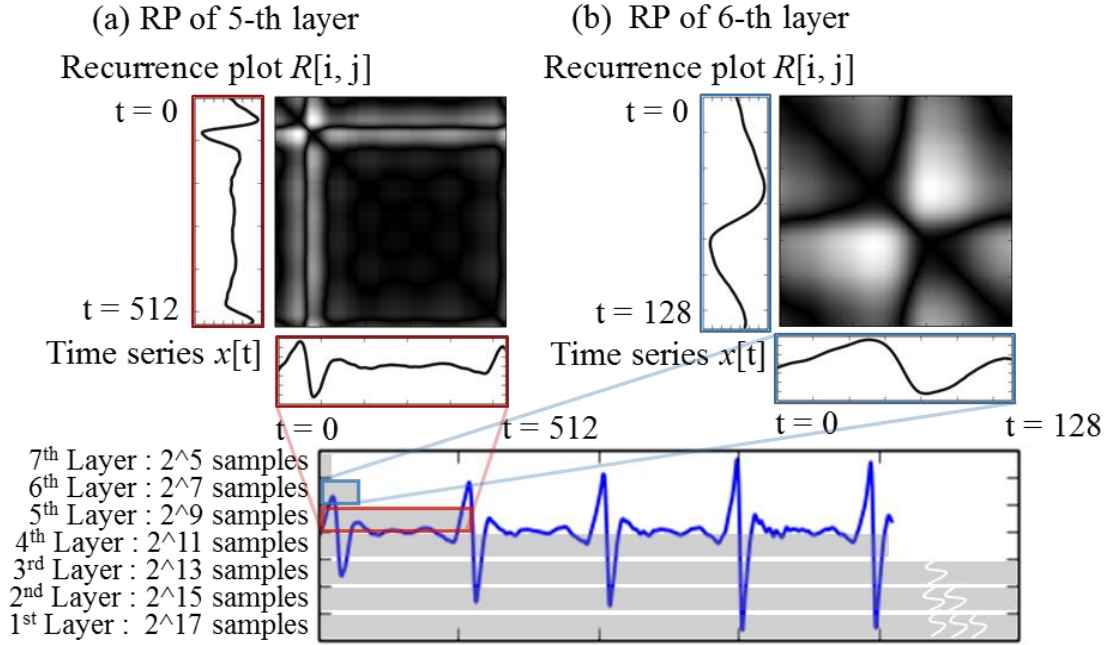

**Fig. 1.** Dimension information of the proposed MRPs.

**2.2. Image reduction through 1-D and 2-D max-pooling**

Because the final MRP images are to be $2^5$ by $2^5$ in size, a proper image resizing technique should be involved to obtain an effective MRP image matrix. The scheme for reducing an MRP image was designed to fit into the subsampling (max-pooling) process in CNNs. By reducing an image using a max-pooling process, the subsampling process in CNNs can form an image in the same way as an image reduction algorithm used for input image construction. Moreover, the max-pooling of an MRP image maintains the abrupt temporal changes in the time-series such that the MRP preserves the salient points even when reduced in size. Not surprisingly, the max-pooling of MRP images also improves the classification performance compared to an average pooling method. The subsampling process for an MRP is described in figure 2. The proposed subsampling process has two phases. The first phase is 1-D max pooling, which simply pools out the maximum value from the input time-series through the equation below:

$$t_{max} = \arg\max\left(|x[t]|, |x[t-1]|, \cdots, |x[t-(N-1)]|\right)$$
$$x_{max1D} = x[t_{max}]$$
(2)

where $N$ is the max pooling size. Unlike the 2-D max pooling process for an image, we preserve the polarity of a time-series sample because an RP computes the difference between two sample positions. After applying a 1-D max pooling process, for the second phase, we construct an MRP image and reduce it in size use 2-D max pooling.

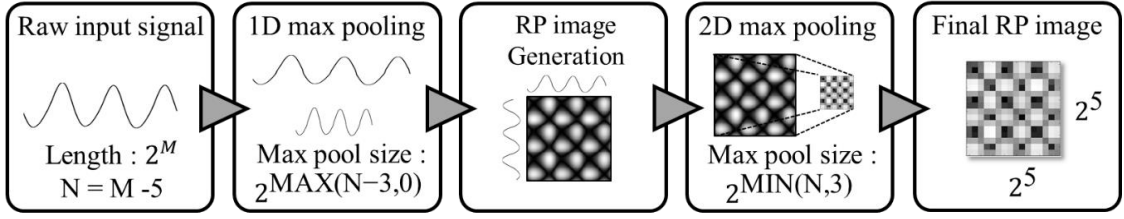

**Fig. 2.** Proposed subsampling process for an MRP image.

**2.3. Data preprocessing**

The magnitude of an MRP image is compressed using the square-root function to reduce the dynamic range of the input signal. This process can be viewed as Steven's power law [16], which reveals the numerical relationship between the magnitude of a physical stimulus and the intensity perceived by humans. A similar concept was also introduced in a Perceptual Linear Predictive (PLP) analysis of speech [17]. Another preprocessing technique is the zero centering of an RP image, which balances negative and positive sample data. However, unlike conventional image processing tasks, we do not normalize the magnitude of an MRP image because the magnitude reveals information regarding the attack and transient shape of a musical tone. Thus, the entire preprocessing technique can be summarized through the following notation, where the square-root and mean functions are element-wise operations.

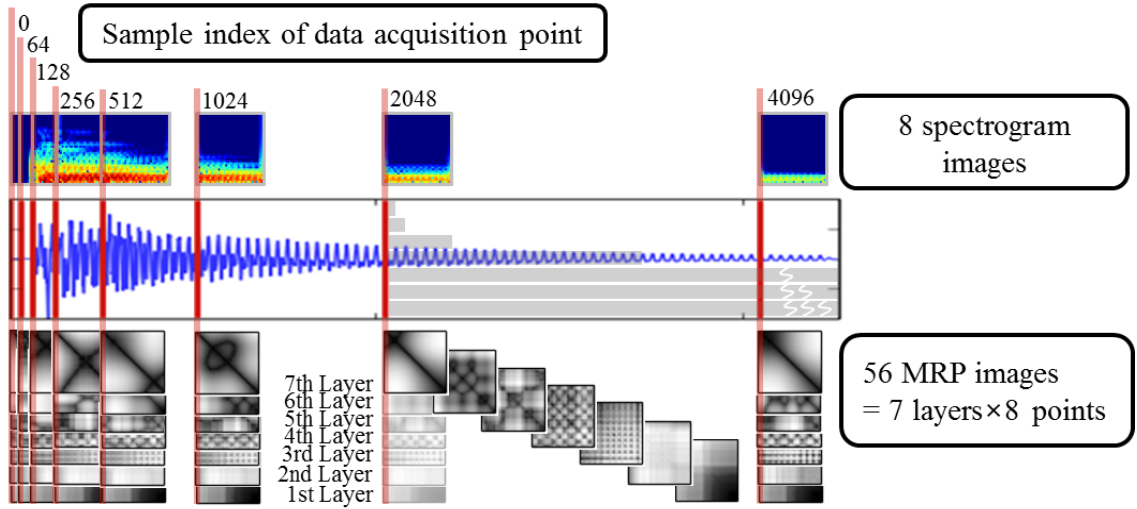

**Fig. 3.** Temporal data acquisition point.

$$\hat{R} = \sqrt{R} - mean\left(\sqrt{R}\right) \tag{3}$$

## 3. NETWORK STRUCTURE AND DATA FEEDING

### 3.1. Data augmentation and temporal acquisition point

Because the onset is one of the most irregular parts of a musical tone in that it contains the unique characteristics of musical instrument sound signal, we concentrated the data acquisition points at the beginning of the signal and sparsely positioned at the rear part of the signal. Eight temporal data acquisition points for a single instrument sample signal are shown in figure 3.

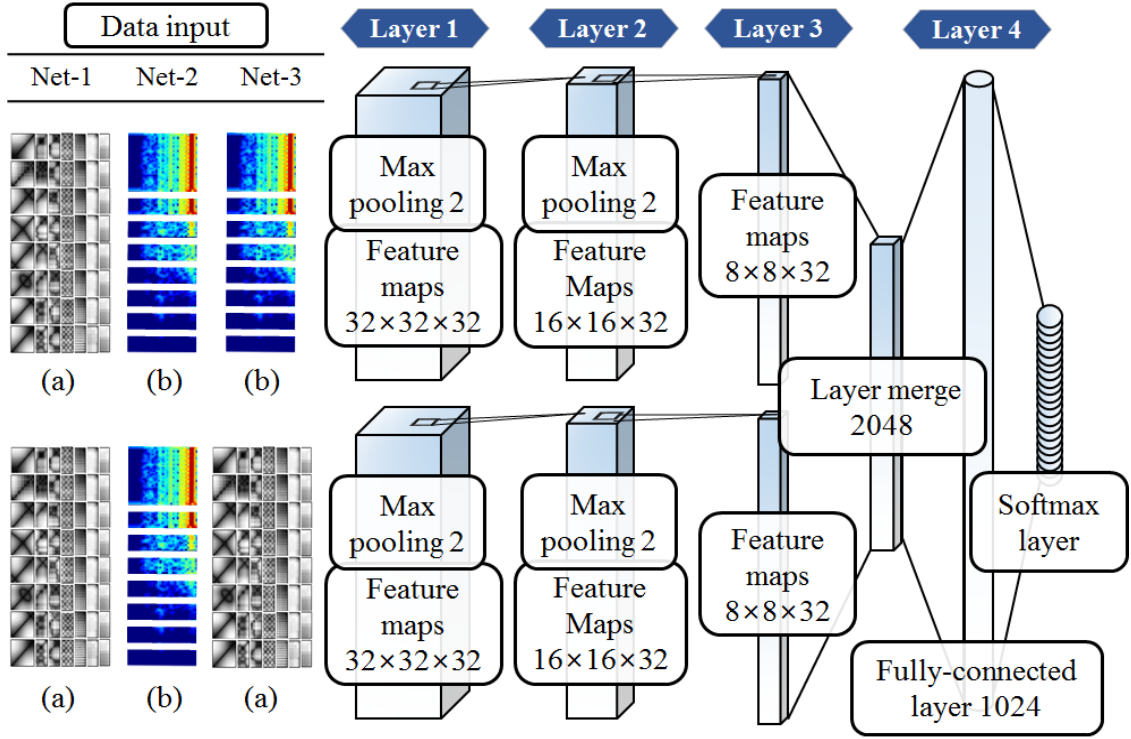

**Fig. 4.** The proposed network structure combining two different data sources. (a) 56 MRP images (b) 8 spectrogram images

To increase the performance of our classifier with data augmentation, we also shifted a time block with 13 samples and made 13 strides to create 13 temporally shifted MRP datasets. For layers of the MRPs, we constructed seven layers of RP images with a quadrupled length of the time-series block compared to the previous block. The temporal sizes of the layers are $2^5$, $2^7$, $2^9$, $2^{11}$, $2^{13}$, $2^{15}$, and $2^{17}$. Thus, we built a total of 56 matrices, with each one containing 32 x 32 MRP images, as described in figure 3.

### 3.2. Data feeding into CNNs and the effect of MRPs

Typical CNNs for image recognition generally employs three channels, red, green and blue, whereas CNNs designed for video classification [18] employs multiple channels for the temporal extent while also utilizing three color channels. Compared with CNNs for video classification, the multiresolution layers can be likened to color channels, and the temporal stride in a times-series can be likened to the temporal extent in video classification.

The merit of using MRPs comes from their multiresolution layers because musical instrument sounds with neighboring notes usually share a similar timbre and phase structure. Because all of the kernel-convoluted images for each channel in a CNN blend into every feature map in the first hidden layer, the CNNs can share kernels for the input of the multiple layers of the MRPs. This makes it easier for the CNNs to learn the features from the neighboring notes in a training set and recognize a test set sample based on the neighboring notes of the training set samples.

Spectrogram image sources were extracted at the same temporal point as that of an MRP. We also used a 32 x 32 spectrogram image with a window of 64 samples and a 1/4 hop length. We employed a linear frequency scale and the magnitude was compressed using the log scale. Both the MRPs and spectrogram images were processed at a 44.1-kHz sampling rate, which corresponds to the sampling rate of the data source.

### 3.3. Network structure

The basic structure of the proposed system is based on the CNNs presented in [7]. However, to integrate two data sources effectively, we employed a multi-column CNNs structure that combines two different convolutional layers from different networks. Multi-column approaches for CNNs have been investigated in several works [18, 19], each of which employs its own method of merging multiple streams of processing. In the case of our system, we merged two streams of processing at the beginning of the fully connected layer level, which adds up the output value of two different convolutional layers, as shown in figure 4. We used a kernel size of 3 x 3, a rectified linear unit (RELU) activation function [20], a stochastic gradient descent (SGD) optimizer (learning rate $1 \cdot 10^{-2}$, decay $1 \cdot 10^{-4}$, momentum 0.8) [21], and a drop out [22] factor of 0.25 for layers 1 and 2, and 0.5 for layers 3 and 4.

| Class Error Rates | task-1 | task-2 | task-3 |
|---|---|---|---|
| Baseline [5] | 3.00% | 13.10% | |
| MRP Network-1 | 2.72% | 12.03% | 13.20% |
| Spectrogram Network-2 | 1.79% | 7.47% | 13.36% |
| Combined Network-3 | **1.31%** | **6.35%** | **6.86%** |

**Table 1.** Classification error rates of all tasks.

## 4. EXPERIMENTAL RESULTS

### 4.1. Test datasets

To evaluate the performance of our classifier system, we used the UIOWA MIS database [23] and piano samples from Virtual Studio Technology (VST) instruments. First, the UIOWA MIS database, which has been widely used in other works [2, 5, 24, 25], contains recordings of single tones of many different musical instrument sounds such as pianos, woodwinds, and brass and string instruments. A subset of 20 musical instruments belonging to four instrument families was selected from the UIOWA MIS database, which is identical to a previous work [5]. Because the classification system in [5] explored a huge variety of handcrafted feature sets and showed the highest classification performance among those works that include the UIOWA MIS database, we set [5] as a baseline for our investigation. To make an exact comparison, we followed the same experimental conditions in [5], i.e., a ten-fold cross validation, and the same instrument family category, instrument labels, and number of samples. We classified four different instrument families of 20 different instruments (task-1) and classified each instrument (task-2). Figure 5 shows the classified instruments and instrument family categories based on the confusion matrix of task-2 using network-3.

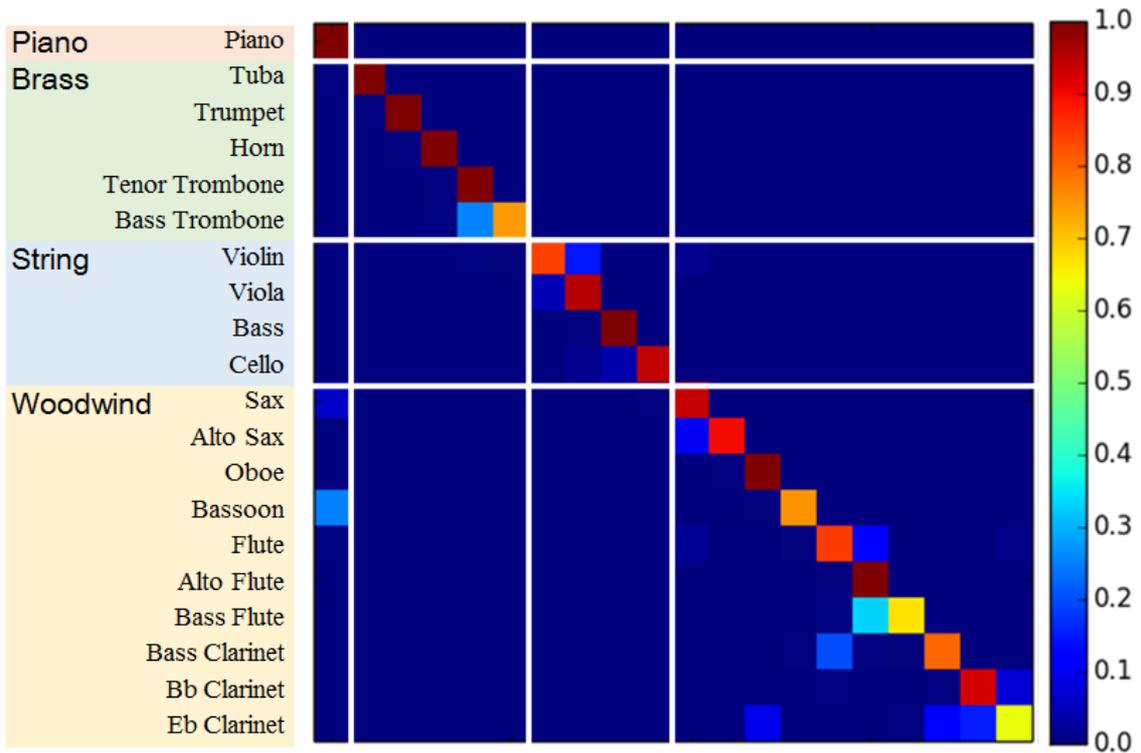

**Fig. 5.** Confusion matrix of classification task-2 with Network-3.

In addition, to evaluate the performance of a more challenging timbre classification for musical instruments, we set up our own dataset classifying four different kinds of pianos using a single note (task-3).

We recorded four seconds of 88 single notes from each of the four different pianos with VST instrument samples (large grand piano, small grand piano, large upright piano, and small upright piano), which recorded the individual notes of the physical pianos without any audio compression or pitch shifting. We also used the same ten-fold cross validation as in task-1 and task-2.

**4.2. Classification performance and discussion**

In table 1, we listed the performances of the MRP-based classification (network-1), spectrogram-image based classification (network-2), and combined the results using multi-column CNNs (network-3) to show the improvement from incorporating MRPs with spectrogram image data. To level the playing field, network-1 and network-2 were designed using two merged networks, the same as network-3, and fed the same data for each sub-network. In addition, we set three networks with identical hyper parameters as we shown in figure 4 and section 3.3. The only factor that was changed is the number of channels for MRP-based networks, and their following number of kernels per channel. For the number of epochs used for the training, we stopped at ten for task-1 and 30 for task-2 and task-3.

For task-1 and task-2, using only network-1, the classification performance was higher than the baseline. However, spectrogram-based network-2 showed an even more improved performance than network-1, even though it observed shorter temporal span than MRPs. Finally, using the proposed multi-column CNNs, we obtained a lower average error rate compared to a network based solely on a spectrogram. For task-3, network-3 also showed an improved performance compared to the other two networks. In task-3, compared with task1 and task-2, the performance of network-2 was not superb, unlike what we saw through task-1 and task-2. This is supposedly caused by the similarity of the temporal envelope of the piano tones. From comparing the results of task-3 with those of task-1 and task-2, we were able to realize that the spectrogram method is very good at analyzing the spectral envelope of each frequency, whereas MRPs are specialized at analyzing the internal phase structure of the instrument tone. Moreover, the improvement of network-3 can be interpreted as the integration of information from different feature domains improving the classification performance.

## 5. SUMMARY

In this paper, we proposed a system for classifying musical instrument sounds. For the instrument classifier, we proposed a new method for extracting the phase trajectory of an acoustic signal using MRPs. Using this proposed technique integrated with spectrogram images, we were able to obtain an improved instrument classification performance using the UIOWA MIS database. In addition, we also tested the performance of our combined network for a piano classification task. The results of the proposed scheme suggest that the classification of musical instrument timbre can be improved using MRP data with a spectrogram image, and by feeding the data and image into multi-column CNNs.